\documentclass[12pt]{article}

\parskip=\medskipamount            
\def\7#1#2{\mathop{\null#2}\limits^{#1}}        
\def\ast{\displaystyle *}

\def\beee{\begin{equation}}
\def\eeee{\end{equation}}
\def\dggg{^{\dagger}}

\begin{document}
\bibliographystyle{unsrt}

\begin{center}
\textbf{REVIEW OF THE\\ N-QUANTUM APPROACH TO BOUND STATES}\\
\vspace{5mm}
O.W. Greenberg\footnote{email address, owgreen@umd.edu}\\
{\it Center for Fundamental Physics\\
Department of Physics \\
University of Maryland\\
College Park, MD~~20742-4111, US}\\
and {\it Helsinki Institute of Physics\\
P.O.Box 64\\
FIN-00014 University of Helsinki\\
Finland}

University of Maryland Preprint PP-10021\\
\end{center}

\begin{abstract} 
     
We describe a method of solving quantum field theories using operator
techniques based on the expansion of interacting fields in terms of asymptotic fields.
For bound states, we introduce an asymptotic field for each (stable) bound state.
We choose the nonrelativistic hydrogen atom as an example to illustrate the
method. Future work will apply this N-quantum approach to relativistic theories that include bound states
in motion.

\end{abstract}

\section{Introduction}

We describe a method for solving of quantum field
theories (QFT) that has wide applicability, and is analogous to methods developed
to solve problems in
classical theories. To find approximate solutions of a theories with a given
Lagrangian, the method approximates the Lagrangian fields by a series of 
normal-ordered asymptotic fields that terminate with a product of up to 
$N$ asymptotic fields. This $N$ is the ``N'' of the N-quantum
approximation. R. Haag introduced this expansion~\cite{haa}. We modified his work by  
introducing asymptotic fields for the stable bound states, if any,
in the theory. We derive analogs of the Schr\"odinger equation that have 
solutions when the theory has bound states.
 
We apply this method to the hydrogen atom
to illustrate how this works for bound states.
Following C.N. Yang and D. Feldman~\cite{yan}, 
we use the QFT
analog of the solution of classical equations of motion in which the the classical 
equations govern a set of functions while the QFT equations
of motion govern a set of operators. As stated above we take account of the operator aspect of
quantum field theory by using a complete and irreducible set of asymptotic fields,
either the \textit{in} or the \textit{out} fields~\cite{haa}.  
A complete set of fields is one that creates
all the states of the Hilbert space by acting repeatedly on a given vector, often
the vacuum, called the cyclic vector.  (More accurately, the complete set needs only to
approximate any vector when polynomials in the operators, integrated
(``smeared'') using smooth functions, act on the cyclic vector.)  An irreducible
set of fields is a set that has no proper invariant subspace. 
By Schur's lemma, any operator that commutes with all operators in an irreducible
set must be a multiple of the identity operator.  Saying that the \textit{in} (or \textit{out})
fields comprise a complete (and irreducible) set, means that any operator in the theory
can be expressed in terms of them.

The \textit{in} (or \textit{out}) fields have free-field commutators, obey free equations of motion,
and the different \textit{in} (or \textit{out}) fields commute or anticommute with each other everywhere
in spacetime.  Each of these sets of asymptotic fields is completely
known once the masses, spins and quantum numbers of the fields in a given set are 
defined.  Thus, either set serves as a collection of standard building blocks from 
which we can construct solutions of the operator equations of motion.  

To solve the equations of motion using the Yang-Feldman approach, we
expand the fields that
appear in the Hamiltonian or Lagrangian in normal-ordered series of \textit{in} (or \textit{out})
fields. (This is the Haag expansion.) To determine the $c$-number amplitudes (the Haag amplitudes) 
that are the coefficients of the normal-ordered
terms, we insert the expansion in the operator equations of motion, again normal-order the asymptotic fields,
and equate the coefficients of corresponding (linearly independent)
normal-ordered terms.  

In this paper, we illustrate this technique for the
case of the nonrelativistic hydrogen atom~\cite{ray}.   The amplitudes for the
bound states satisfy the Schr\"odinger equation for the hydrogen atom and, of course, we
find the usual spectrum of bound states.  This shows that the relevant Haag 
amplitudes are the wave functions of the bound states.  

Our method, which is based on the Haag expansion, 
is entirely independent of the Bethe-Salpeter equation. 
In contrast to the Bethe-Salpeter approach, our method 
contains no spurious solutions and no negative norm
amplitudes.  The method can be used for bound states in relativistic theories,
despite that the amplitudes depend on the same number of kinematic
variables as appear in nonrelativistic wave functions. There are no
relative time intervals in the relativistic version of this method.

We applied this formulation previously to a model of the deuteron~\cite{nqab} in an
approximation where this method is similar to that of the Gross equation~\cite{gro}, but differs
from it in higher-order approximations.

In this paper, we neglect two related issues that are relevant for the Coulomb interaction. 
One is that, because the Coulomb interaction is long ranged, spatially separated
particles never move independently~\cite{got}. This does affect any questions addressed in our study. 
The other point is that the Coulomb field of charged particles is not
attached to the particles, as suggested by P.A.M. Dirac~\cite{dir}. Nevertheless, this too does not appear 
to cause any difficulty for the conclusions drawn from this work.

Although our method does not require expressions for interpolating (Heisenberg) 
$in$ and $out$ asymptotic fields for bound states,
we show how to construct such interpolating fields for the bound states of the theory
(in our case, the states of the hydrogen atom) in terms of the fundamental fields that
enter the Hamiltonian of the theory. This construction, which uses a convolution of
the bound-state
wave functions with the constituent fields, 
contrasts with the constructions of K. Nishijima~\cite{nish} 
and of W. Zimmermann~\cite{zimm}, which
use products of the fundamental fields at the same point.

We plan to apply the construction of interpolating fields for bound states 
of baryons and mesons to calculations based on quark models within quantum chromodynamics.

We note that when we use $in$ ($out$) fields as the asymptotic fields, we have 
retarded (advanced) boundary conditions on the amplitudes (Green's functions) that 
provide dispersion relations for these amplitudes, which G. K\"{a}ll\'{e}n~\cite{kal}
discussed in the context of perturbation theory for Yang-Feldman equations with retarded boundary conditions.

\section{Asymptotic fields and the Haag expansion}

Asymptotic fields have been part of quantum field theory at least since
the work of H. Lehmann, K. Symanzik and W. Zimmermann~\cite{leh}; nevertheless, there are
still misconceptions that should be cleared up.  Asymptotic fields are free
fields in theories that have no long-range potentials, massless particles
or confinement, because, in such theories, we describe particles either by:
(i) separated 
asymptotic fields (at large time scales) or (ii) as
a bound state, also represented by an independent
asymptotic field.  In either case, the exact eigenstates can be labeled by the quantum
numbers of the free particles.  The (weak) limits for $t \rightarrow \pm \infty$ are the
{\it out} or {\it in} fields that construct the eigenstates at the corresponding 
limiting times~\cite{thesis}.  The nontrivial unitary relation
between these fields is given by the ${\it S}$-operator,
\beee
S \phi^{out}(x) S\dggg=\phi^{in}(x).
\eeee
The asymptotic fields at finite times are the limiting fields
propagated to finite times according to the free equations of motion.
The free-field property of the asymptotic fields does not depend
on any unphysical ``adiabatic switching off'' of interactions.
Even when the potentials are long-range (and there are massless 
particles in the context of the field theory), we can use often
use the usual asymptotic fields. as is the case for
the example considered in this paper. 

The limits that define the asymptotic fields are subtle.  The relations that
appear in some books,
\beee
A(\mathbf{x},t) \rightarrow A^{(out)in}(\mathbf{x},t), ~~t \rightarrow \pm \infty,
\eeee
are ill-defined. We discuss the proper (weak-limit) definition of the asymptotic fields in 
the Appendix.

\section{Nonrelativistic model of the hydrogen atom}

The fundamental fields are the spin 1/2 electron, $e_{a}(\mathbf{x},t)$, and the spin 1/2 proton, 
$p_{a}(\mathbf{x},t)$, fields. The $a$ indices label the spin components of the
fields. We assume that these fields obey 
the usual equal-time canonical anticommutation relations,
\beee
[e_{a}(\mathbf{x},t),e_{b}\dggg(\mathbf{y},t)]_+=\delta_{ab}\delta(\mathbf{x}- \mathbf{y}),
\eeee
\beee
[p_{a}(\mathbf{x},t),p_{b}\dggg(\mathbf{y},t)]_+=\delta_{ab}\delta(\mathbf{x}- \mathbf{y}),
\eeee
and that the other equal-time anticommutators vanish.  The nonrelativistic Hamiltonian is 
\begin{eqnarray}
H & = & \frac{1}{2m}\int d^3x \nabla_{\mathbf{x}}e_{a}\dggg(\mathbf{x},t)\cdot 
\nabla_{\mathbf{x}}e_{a}(\mathbf{x},t)
+ \frac{1}{2M}\int d^3x \nabla_{\mathbf{x}}p_{a}\dggg(\mathbf{x},t)\cdot 
\nabla_{\mathbf{x}}p_{a}(\mathbf{x},t)        \nonumber        \\
& & -\int d^3x d^3y p_{b}\dggg(\mathbf{y},t)e_{a}\dggg(\mathbf{x},t) \frac{e^2}{|\mathbf{x}-\mathbf{y}|}
e_{a}(\mathbf{x},t)p_{b}(\mathbf{y},t),
\end{eqnarray}
where $m$ and $M$ are the electron and proton masses.  (We have not included spin-dependent terms in
the interaction.) The equation of motion for the
electron field is
\beee 
i \partial_t e_{a}(\mathbf{x},t)=-\frac{1}{2m} \nabla^2_\mathbf{x} e_{a}(\mathbf{x},t)
-\int d^3y p_{b}\dggg(\mathbf{y},t)
p_{b}(\mathbf{y},t)\frac{e^2}{|\mathbf{x}-\mathbf{y}|}e_{a}(\mathbf{x},t).
\eeee
The proton field obeys a similar equation.

\section{Derivation of the Schr\"odinger equation for the hydrogen atom}

We illustrate the N-quantum approach for the hydrogen atom in the approximation
in which the excited states of hydrogen are stable.  We introduce the terms
for this case into the Haag expansions of the electron and proton fields. The
relevant \textit{in} fields (alternatively, the \textit{out} fields can be used as well) 
are those for the electron, $e_{a}^{in}(\mathbf{x},t)$, the proton, $p_{a}^{in}(\mathbf{x},t)$
and the \textit{in} fields for each state of the hydrogen atom, $h^{in}_{n^{\prime}l^{\prime}m^{\prime}}$.  
For simplicity,
we choose the $F=0$ state of the hydrogen atom, so the $in$ field, $h^{in}_{n^{\prime}l^{\prime}m^{\prime}}$,
is a scalar boson.
(We use the known spectrum of the hydrogen atom. In general, we would use
the Schr\"odinger equation, which we derive below, to find any such bound states.) The
\textit{in} fields obey free equations of motion and free anticommutation (commutation) relations:
For the electron, we have,
\beee
i \partial_t e_{a}^{in}(\mathbf{x},t)=-\frac{1}{2m} \nabla^2_\mathbf{x} e_{a}^{in}(\mathbf{x},t),
\eeee
\beee
[e_{a}^{in}(\mathbf{x},t),e_{b}^{in \dagger}(\mathbf{y},t^{\prime})]_+
=\delta_{ab}{\cal D}(\mathbf{x}-\mathbf{y},t-t^{\prime};0,m),
\eeee
\beee
{\cal D}(\mathbf{x},t;E,m)=\frac{1}{(2 \pi)^3}\int d \omega d^3k 
\delta(\omega-E-\frac{{\mathbf k}^2}{2m})e^{-i \omega t+i {\mathbf k} \cdot \mathbf{x}}.    \label{d}
\eeee
The proton $in$ field obeys the analogous equations, with the proton mass replacing the
electron mass.  The $in$ fields for the states of the hydrogen atom obey,
\beee
i \partial_t h^{in}_{nlm}(\mathbf{x},t)=(E_{nlm}-\frac{1}{2M_{nlm}} \nabla^2_\mathbf{x})
h^{in}_{nlm}(\mathbf{x},t),
\eeee
\beee
[h^{in}_{nlm}(\mathbf{x},t),h^{in \dagger}_{n^{\prime}l^{\prime}m^{\prime}}(\mathbf{y},t^{\prime})]_-
=\delta_{nlm,n^{\prime}l^{\prime}m^{\prime}}
{\cal D}(\mathbf{x}-\mathbf{y},t-t^{\prime};E_{nlm},M_{nlm}),
\eeee
where $M_{nlm}=m+M+E_{nlm}$ is the mass of the hydrogen atom  in the state $nlm$. 
The anticommutators or commutators between different $in$ fields vanish.

Because the electron is stable the appropriate terms in the Haag expansion for the electron field are the electron $in$
field, and for the states of the hydrogen atom, the hydrogen $in$ fields.
In the annihilation part for the electron, the
operators that enter the bound-state term are the hydrogen annihilation operator and the
creation operator for the proton.  The quantum numbers balance, in that, for example, both
$e^{in}$ and $p^{in \dagger}h^{in}_{nlm}$ annihilate charge $-e$ and lepton number $+1$. 
(For simplicity, we choose the $F=0$ state of the hydrogen atom, so that $h^{in}_{nlm}$ represents a 
scalar boson.)
As will be shown explicitly below, this term in the Haag amplitudes corresponds to the usual 
bound-state wave functions for the hydrogen atom. 
We choose a scalar hydrogen atom to simplify our discussion. Using the constraint~\cite{cor} from the nonrelativistic
boost, $\mathbf{k} \rightarrow \mathbf{k} + \mathbf{v}$, we find
\begin{eqnarray}
e_{a}(\mathbf{x},t) &=& e^{in}_{a}(\mathbf{x},t)      \nonumber \\                     
& & + \sum_{nlm} \int \psi^e_{nlm}(\mathbf{x} - \mathbf{y})
\epsilon_{ab}:p_b^{in \dagger}(\mathbf{y},t)
h^{in}_{nlm}(\frac{m \mathbf{x}+M\mathbf{y}}{m+M},t):d^3y,   \label{sch}
\end{eqnarray}
The analogous expansion holds for the proton field:
\begin{eqnarray}
p_a(\mathbf{y},t) & = & p_a^{in}(\mathbf{y},t)    \nonumber  \\                  
& & + \sum_{nlm} \int \psi^p_{nlm}(\mathbf{x} - \mathbf{y})
\epsilon_{ab}:e_b^{in \dagger}(\mathbf{y},t)
h^{in}_{nlm}(\frac{m \mathbf{x}+M\mathbf{y}}{m+M},t):d^3y.
\end{eqnarray}
The wave function $\psi^e_{nlm}$ is the matrix element of the Hamiltonian field
$e_{a}(\mathbf{x},t)$ between the proton in state and the in states for the hydrogen atom:
\beee
_{in}\langle p_b(\mathbf{y},t)|e_{a}(\mathbf{x},t)|h(\mathbf{\mathbf{X}},t)\rangle_{in}
=\epsilon_{ab}\psi^e_{nlm}(\mathbf{x} - \mathbf{y})\delta(\frac{m \mathbf{x}+M\mathbf{y}}{m+M}-\mathbf{X}).
\eeee

To find the equation for the wave function $\psi^e_{nlm}$, we substitute these expansions into
the equation of motion for the electron field, again normal-order the $in$ fields, and 
examine the terms that contain $p_b^{in \dagger}h^{in}_{nlm}$:
\begin{eqnarray}
i \partial_t \int \sum_{nlm}
\psi^e_{nlm}(\mathbf{x} - \mathbf{y},\mathbf{x}-\mathbf{X}) \epsilon_{ab}:p_b^{in \dagger}(\mathbf{y},t)
h^{in}_{nlm}(\frac{m \mathbf{x}+M\mathbf{y}}{m+M},t):d^3y  & = & \nonumber  \\
\sum_{nlm}\int (-\frac{1}{2m}
\nabla^2_{\mathbf{x}} -\frac{e^2}{|\mathbf{x}-\mathbf{y}|}) 
 \psi^e_{nlm}(\mathbf{x} - \mathbf{y}) 
 \epsilon_{ab} :p_b^{in \dagger}(\mathbf{y},t) h^{in}_{nlm}(\frac{m \mathbf{x}+M\mathbf{y}}{m+M},t):d^3y.
\end{eqnarray}
To find the Schr\"odinger equation for the hydrogen atom with momentum $\mathbf{P}$, we 
introduce the Fourier transform: 
\beee
h^{in}_{nlm}(\mathbf{x},t)=\frac{1}{(2 \pi)^{3/2}} 
\int \tilde{h}^{in}(\mathbf{P}) exp(-i(E_{nlm}+ \frac{\mathbf{P}^2}{2 M_{nlm}}) t+i\mathbf{P} \cdot \mathbf{x})d^3P.
\eeee
In terms of the Fourier transform, the equation is
\begin{eqnarray}
\sum_{nlm} \int 
\psi^e_{nlm}(\mathbf{x} - \mathbf{y}) \epsilon_{ab}(i \partial_t + E_{nlm} + \frac{\mathbf{P}^2}{2 M_{nlm}})
:p_b^{in \dagger}(\mathbf{y},t)
\tilde{h}^{in}_{nlm}(\mathbf{P}):d^3y  & = & \nonumber  \\
\sum_{nlm}\int (-\frac{1}{2m}
\nabla^2_{\mathbf{x}} -\frac{e^2}{|\mathbf{x}-\mathbf{y}|}) 
 \psi^e_{nlm}(\mathbf{x} - \mathbf{y}) 
 \epsilon_{ab} :p_b^{in \dagger}(\mathbf{y},t) \tilde{h}^{in}_{nlm}(\mathbf{P}):d^3y.
\end{eqnarray}

We evaluate the time derivative on the left hand side using the equation for the
$p_b^{in \dagger}$, and integrate by parts to get derivatives acting on the wave function. 
The left hand side becomes
\[
\sum_{nlm} \int (\frac{1}{2M}\nabla_\mathbf{y}^2 + E_{nlm}
+\frac{\mathbf{P}^2}{2M_{nlm}})\psi^e_{nlm}(\mathbf{x} - \mathbf{y})
\epsilon_{ab}:p_b^{in \dagger}(\mathbf{y},t))\tilde{h}^{in}_{nlm}(\mathbf{P}):d^3y. 
\]
We do not assume the hydrogen atom is at rest; the last term above gives its kinetic energy.
We are using the kinematics for the nonrelativistic
limit of a Lorentz covariant theory. (For a Galilean covariant theory, the mass of the
hydrogen atom is $m+M$ for all states of the hydrogen atom because of the Bargmann
superselection rule.~\cite{gal})
The final step is to 
equate the coefficients of the $\:p_b^{in \dagger}\tilde{h}^{in}_{nlm}\:$ terms.  We justify this either
by noting that these terms are linearly independent of all other types of $in$ field products
or, more formally, by anticommuting with $p_b^{in}$ and commuting with $h^{in}_{nlm}$.  
The result is
\beee
(-\frac{1}{2m} \nabla_\mathbf{x}^2 -\frac{1}{2M}\nabla_\mathbf{y}^2 -\frac{e^2}{|\mathbf{x}-\mathbf{y}|})
\psi^e_{nlm}(\mathbf{x}-\mathbf{y}) = (E_{nlm}+\frac{\mathbf{P}^2}{2M_{nlm}})
\psi^e_{nlm}(\mathbf{x}-\mathbf{y})),
\eeee
\beee
E_{nlm}=-\frac{mMe^4}{2(m+M)n^2},
\eeee
where we used the known energy levels of the hydrogen atom. 
We recognize this as the Schr\"odinger equation for the hydrogen atom and identify the
$\psi^e_{nlm}$ as its wave functions.

In terms of the relative coordinate, $\mathbf{r}=\mathbf{x}-\mathbf{y}$, the Schr\"odinger
equation for the
hydrogen atom with momentum $\mathbf{P}$ is
\beee
(-\frac{1}{\mu_{nlm}}\nabla_{\mathbf{r}}^2-\frac{e^2}{|\mathbf{r}|})\psi^e_{nlm}(\mathbf{r})
=(E_{nlm} + \frac{1}{\mu_{nlm}}\mathbf{P}^2)\psi^e_{nlm}(\mathbf{r}),
\eeee
where $1/\mu=1/m + 1/M$ and $1/\mu_{nlm}=1/M_{nlm}-1/(m+M)$.

Because $1/\mu_{nlm} > 0$, the relative wave function for the bound state in motion obeys
an equation with a (slightly) higher energy than that for the bound state at rest.
In a Galilean invariant theory, $M_{nlm}=M+m$ for all states of the hydrogen atom, and the 
atom in motion obeys the same equation as the atom at rest because of the Bargmann
mass superselection rule~\cite{gal}.

The bound states of the hydrogen atom, taking into account the spin of the electron and the proton, 
are labeled by $nLSJJ_z$, and the N quantum amplitudes also have these
labels, as well as spin indices $a,b$ for the electron and proton spins. For the $S=0$
states, the amplitude is
\beee
\psi^{e~J}_{ab~J_z}(nL~S=0)(\mathbf{r})=\epsilon_{ab} 
\psi^{e~J}_{J_z}(nL~S=0)(\mathbf{r}).
\eeee
For the $S=1$ states the amplitude is
\beee
\psi^{e~J}_{ab~J_z}(nL~S=1)(\mathbf{r})=\sigma^{S_z}_{ab} 
\psi^{e~J}_{J_z}(nL~S=1)(\mathbf{r}),
\eeee
where $J=L-1,L,L+1$.
The corresponding annihilation operators for the $S=0$ hydrogen atom are
\beee
h^{J}_{J_z}(nLS,J=L;\mathbf{R})=h^{LS=0}_{L_z,J_z=L_z}(\mathbf{R})
\eeee 
and for $S=1$, 
\begin{eqnarray}
\lefteqn{h^{J}_{J_z}(nLS,J=L-1,L,L+1;\mathbf{R})=} \nonumber \\
& & \sum_{L_z}\sqrt{2J+1}(-)^{L-S-J_z}\left( \begin{array}{ccc}
L & S & J  \\
L_z &  J_z-L_z & J_z
\end{array} \right)
h^{LS}_{L_z,J_z-L_z=S_z}(\mathbf{R}).
\end{eqnarray}

\section{Normalization of the wave functions}

The asymptotic fields diagonalize conserved observables such as the 
Hamiltonian, the momentum operators and various charges.  Thus, any such conserved
quantity has the form of a sum of free operators, ${\cal O}_{free}\{\phi_{in}\}$, for the 
contribution of
each $in$ field, including those for bound states, to the observable ${\cal O}$.  
For the $\psi^e_{nlm}$,
with the electron off-shell, we choose the lepton number $L$,
whose term in the $in$ field expansion provides the normalization condition, to be the observable. The
lepton number in this model is
\beee
L=\int d^3x e\dggg(\mathbf{x},0)e(\mathbf{x},0)=L_0\{e^{in}\}+\sum_{nlm}L_0\{h^{in}_{nlm}\}+ \cdots.
\eeee
We substitute the Haag expansion for $e(\mathbf{x},t)$ given in Eq.(\ref{sch}) into $L$, look for the
terms bilinear in $h^{in}_{nlm}$, and equate these to 
$L_0\{h^{in}_{nlm}\}=\int d^3R h^{in \dagger}_{nlm} h^{in}_{nlm}$.  We find
\begin{eqnarray}
\lefteqn{\int d^3x \sum_{nlm} \psi^{e \ast}_{nlm}(\mathbf{x} - \mathbf{y}) 
h^{in \dagger}_{nlm}({\mathbf R},0)p^{in}(\mathbf{y},0)d^3y}      \nonumber \\        
& & \times \sum_{n^{\prime}l^{\prime}m^{\prime}} 
\psi^e_{n^{\prime}l^{\prime}m^{\prime}}(\mathbf{x} - {\mathbf y^{\prime}})
p^{in \dagger}({\mathbf y^{\prime}},0) h^{in}_{n^{\prime}l^{\prime}m^{\prime}}
({\mathbf R}^{\prime},0)d^3y^{\prime} =  \nonumber \\                                     
& & \sum_{nlm} \int d^3R h^{in \dagger}_{nlm}({\mathbf R},0) h^{in}_{nlm}({\mathbf R},0) 
\end{eqnarray}
with $R=(m\mathbf{x}+M\mathbf{y})/(m+M) and ~R^{\prime}=(m\mathbf{x}+M\mathbf{y}^{\prime})/(m+M)$.
The $p^{in}$ operators on the left-hand side contract, 
$\langle p^{in}(\mathbf{y},0) p^{in \dagger}({\mathbf y^{\prime}},0) \rangle=\delta(\mathbf{y}-{\mathbf y^{\prime}})$.
When we remove the $h^{in}_{nlm}$ operators by commuting with $h^{in \dagger}_{nlm}$ operators, 
the usual orthonormalization condition results:
\beee
\int d^3r \psi^{e \ast}_{nlm}({\mathbf r}) \psi^e_{n^{\prime} l^{\prime} m^{\prime}}({\mathbf r})
=\delta_{nlm, n^{\prime} l^{\prime} m^{\prime}}.
\eeee

\section{Equal-time anticommutation relations}

The equal-time anticommutation relations give relations among the Haag amplitudes.
These relations follow from the vanishing of the coefficients of each
(linearly independent) normal-ordered product of $in$ fields.  Most of the relations
involve Haag amplitudes for terms with higher-degree normal-ordered products
than we have considered, however, for the equal-time anticommutator $[e,p]_+=0$,
there is a relation that involves the wave functions:
\beee
\psi^e_{nlm}({\mathbf r}) + \psi^p_{nlm}(-{\mathbf r})=0.   \label{eqt}
\eeee
Thus the Haag amplitude for the off-shell electron is related simply to that for
the off-shell proton, and the apparent asymmetry in the treatment of the constituents
of the bound state because one particle is on-shell and one particle
is off-shell is therefore not a true asymmetry.  The two amplitudes determine each other 
uniquely,  and it is convenient to define $\psi_{nlm}({\mathbf r}) \equiv \psi^e_{nlm}({\mathbf r})$.

\section{Definition of asymptotic limits}

The proper asymptotic limit is a weak operator limit that constructs an asymptotic field of a given
mass $m$ from the neighborhood of the mass $m$ part of the relevant (product of)
Lagrangian fields~\cite{thesis}. The asymptotic ($in$ or $out$) fields for (possibly composite)
particles are characterized by their rest energy $E$, mass $m$, and spin $J$. 
Suppressing the spin in what follows, we define the asymptotic fields associated
with the interacting field $A(\mathbf{x}, t)$ as
\beee
A^{in~(out)}(\mathbf{x}, t)=lim_{t^{\prime} \rightarrow \mp \infty} 
\int \mathcal{D}(\mathbf{x}-\mathbf{y}, t-t^{\prime};E_A, m_A)A(\mathbf{y},t^{\prime})d^3y,
\eeee
where the limit is the weak limit of the smeared operators, and
$\mathcal{D}(\mathbf{x}, t)$ is defined in Eq.(\ref{d}).
The asymptotic fields for other interacting fields are defined in an analogous way. 
(We 
define the asymptotic fields for composite particles in Sec. 8.) The asymptotic
limit in momentum space is often useful in calculations:
\beee
\tilde{A}^{in~(out)}(\mathbf{k}, E)=lim_{t^{\prime} \rightarrow \mp \infty} 
\delta(E-\frac{\mathbf{k}^2}{2m})\int dE^{\prime} e^{i(E-E^{\prime})t^{\prime}}
\tilde{A}(\mathbf{k},E^{\prime}).
\eeee
Either form of the definition of the asymptotic limits makes clear that asymptotic
fields obey the free equations of motion:
\beee
i \partial_t A^{in~(out)}(\mathbf{x},t)=(E-\frac{1}{2m} \nabla^2)A^{in~(out)}(\mathbf{x},t),
\eeee
as well as the free field anticommutation or commutation relations:
\beee
[A^{in~(out)}(\mathbf{x},t),A^{in~(out)}(\mathbf{y},t^{\prime})]_{\pm}
=\mathcal{D}(\mathbf{x}-\mathbf{y}, t-t^{\prime}).
\eeee
Note that: 
\beee 
\mathcal{D}(\mathbf{x}, 0)=\delta(\mathbf{x})
\eeee
for all $E,~m$.

\section{Construction of asymptotic fields for bound states}

In this section, we show how to construct the asymptotic fields for the bound
states of the hydrogen atom from a convolution of the wave function with products of electron and
proton fields at separated points. Since the hydrogen atom is made of an electron
and a proton and these fields anticommute, as
an intuitive guess we start with one-half the commutator of 
the electron and proton fields, 
\begin{eqnarray}
\lefteqn{\frac{1}{2}[p(\mathbf{y},t),e(\mathbf{x},t)]_- = }   \nonumber  \\  
& & \frac{1}{2}[p^{in}(\mathbf{y},t) +                             
\sum_{nlm} \int \psi^p_{nlm}(\mathbf{y} - \mathbf{x}^{\prime}),    
e^{in \dagger}(\mathbf{x}^{\prime},t)
h^{in}_{nlm}(\frac{m\mathbf{x}^{\prime}+M\mathbf{y}}{m+M},t)d^3x^{\prime},  \nonumber  \\
& & e^{in}(\mathbf{x},0) +                                
\sum_{n^{\prime}l^{\prime}m^{\prime}} \int \psi^e_{nlm}(\mathbf{x} - {\mathbf y^{\prime}}) 
p^{in \dagger}({\mathbf y^{\prime}},t)
h^{in}_{n^{\prime}l^{\prime}m^{\prime}}
(\frac{m\mathbf{x}+M{\mathbf y^{\prime}}}{m+M})d^3y^{\prime}]_-.  
\end{eqnarray}
After normal-ordering, the terms linear in $h^{in}_{n^{\prime}l^{\prime}m^{\prime}}$ are:
\begin{eqnarray}
\sum_{n^{\prime}l^{\prime}m^{\prime}}\frac{1}{2}
[\psi^e_{n^{\prime}l^{\prime}m^{\prime}}(\mathbf{x} - 
\mathbf{y})-\psi^p_{n^{\prime}l^{\prime}m^{\prime}}(\mathbf{y} - \mathbf{x})]
h^{in}_{n^{\prime}l^{\prime}m^{\prime}}(\frac{m\mathbf{x}+M\mathbf{y}}{m+M}) &= & \nonumber \\
\sum_{n^{\prime}l^{\prime}m^{\prime}}\psi_{n^{\prime}l^{\prime}m^{\prime}}(\mathbf{x} - \mathbf{y})
h^{in}_{n^{\prime}l^{\prime}m^{\prime}}(\frac{m\mathbf{x}+M\mathbf{y}}{m+M}), & & 
\end{eqnarray}
where we used Eq.(\ref{eqt}).
We can use the orthonormality of the hydrogen wave functions to eliminate the
wave functions and isolate the {\it in} field for the states of the hydrogen atom:
The term linear in the hydrogen {\it in} field becomes:
\beee
\int \psi^{\ast}_{nlm}(\mathbf{x} - \mathbf{y})
\psi_{n^{\prime}l^{\prime}m^{\prime}}(\mathbf{x} - \mathbf{y})
h^{in}_{n^{\prime}l^{\prime}m^{\prime}}(\frac{m\mathbf{x}+M\mathbf{y}}{m+M})d^3(\mathbf{x}-\mathbf{y})=
h^{in}_{nlm}(\frac{m\mathbf{x}+M\mathbf{y}}{m+M}).
\eeee
Now we define an interpolating field for the hydrogen
states,
\beee
h_{nlm}({\mathbf R},t)=\frac{1}{2} 
\int d^3r \psi^{\ast}_{nlm}({\mathbf r}) 
[p({\mathbf R}-\frac{m}{m+M}{\mathbf r},t),e({\mathbf R}+\frac{M}{m+M}{\mathbf r},t)]_-.
\eeee
From these calculations, we see that the only term in this interpolating field that is
linear in $h^{in}_{nlm}$ is just $h^{in}_{nlm}$, and further this is the only term with
a singularity at the energy and mass of the bound state.  Therefore this term can be
isolated by taking the weak limit:
\beee
h_{nlm}^{in(out)}({\mathbf R},t)=
lim_{\tau \rightarrow \mp \infty} \int_{t^{\prime}=\tau}
d^3R ~{\cal D}
({\mathbf R}-{\mathbf R}^{\prime},t-t^{\prime};E_{nlm},m+M)h_{nlm}({\mathbf R}^{\prime},t).
\eeee
This discussion of bound states, which involves the bound-state amplitude, 
is an alternative to the constructions of bound-state operators from
products constituent fields at a point that were formulated by
Nishijima~\cite{nish} and by Zimmermann~\cite{zimm}.

\section{Summary}

We derived the Schr\"odinger equation for bound states of
the hydrogen atom using the operator equation of motion for the electron
(or proton) field together with the Haag expansion of the fields in terms
of normal-ordered products of $in$ (or $out$) fields.  This was done
without any reference to
the Bethe-Salpeter equation. The usual normalization conditions and probability
interpretation of bound state amplitudes are valid both in the present
nonrelativistic example as well as for relativistic theories.

The N quantum approach has been applied to many other problems, and a survey of results
appears in Ref.(~\cite{vir}). 

\section{Appendix: the asymptotic limits}

Here we discuss the asymptotic limits of the N-quantum approach,  and indicate how the weak limits
eliminate terms that do not have  delta-function or principal-value
singularities on the energy shell of the asymptotic field.  Considering an operator
${\cal O}(\mathbf{x},t)$, we define the asymptotic limits 
${\cal O}^{in(out)}(\mathbf{x},t)$ by
\beee
{\cal O}^{in(out)}(\mathbf{x},t)=lim_{t^{\prime} \rightarrow \mp \infty}
\int d^3 x^{\prime} {\cal D}(\mathbf{x}-\mathbf{x}^{\prime},t-t^{\prime};E,m)
{\cal O}(\mathbf{x}^{\prime},t^{\prime}),
\eeee

Using a four-dimensional Fourier transform, 
\beee
{\cal O}(\mathbf{x},t)=\frac{1}{(2 \pi)^4}\int d \omega d^3k 
\tilde{{\cal O}}(\omega,{\mathbf k}) e^{-i \omega t+i {\mathbf k} \cdot \mathbf{x}},
\eeee
we find the momentum-space form of the weak asymptotic limit:
\beee
\tilde{{\cal O}}^{in(out)}(\omega,{\mathbf k})=lim_{t^{\prime} \rightarrow \mp \infty}
\delta(\omega - E -\frac{{\mathbf k}^2}{2m})
\int d \omega^{\prime} e^{i(\omega - \omega^{\prime})t^{\prime}} 
\tilde{{\cal O}}(\omega^{\prime},{\mathbf k})
\eeee
As a matrix element between two states, this gives a relation between
distributions.  If the matrix element of $\tilde{{\cal O}}(\omega^{\prime},{\mathbf k})$
on the right hand side is in $L^1$, the 
Riemann-Lebesgue lemma states that the term will vanish in this limit except when 
there is a singularity in the matrix element of 
$\tilde{{\cal O}}(\omega^{\prime},{\mathbf k})$ on the energy shell 
$\omega^{\prime}=E+\frac{{\mathbf k}^2}{2m}$.  The limit is finite if the singularity
is either a delta function in $\omega^{\prime}-E-\frac{{\mathbf k}^2}{2m}$ or has a 
principal value $1/(\omega^{\prime}-E-\frac{{\mathbf k}^2}{2m})$.  The limit
diverges for stronger singularities on the energy shell.  Therefore the requirement that the weak 
asymptotic limit exist eliminates terms that do not have delta-function or principal-value 
singularities on the energy shell.  Ref.(\cite{thesis}) gives further discussion of weak asymptotic
limits.

\section{Acknowledgements}
It is a pleasure to thank Dan-Olof Riska for his hospitality at the Helsinki Institute
for Physics as well as for informative and stimulating discussions. I thank Steven Cowen, Paul Hoyer and Claus 
Montonen for their interest and for their helpful comments. I give special thanks to Tom Ferbel for
many excellent suggestions to improve the readability of this paper.

\end{document}